\documentclass[12pt]{article}
\usepackage[utf8]{inputenc}
\usepackage{a4wide}
\usepackage{amsmath}
\usepackage{amssymb}
\usepackage{amsfonts}
\usepackage{siunitx}
\usepackage{setspace}
\usepackage{verbatim}
\usepackage{rotating}
\usepackage{lscape}
\usepackage{authblk}
\usepackage{color}
\usepackage{bm}
\usepackage{soul}
\newcommand{\ve}{\varepsilon}
\newcommand{\heps}{\widehat{\ve}}
\newcommand{\PIR}{\mbox{PIR}}
\newcommand{\hPIR}{\widehat{\mbox{PIR}}}
\newcommand{\tPIR}{\widetilde{\mbox{PIR}}}
\newcommand{\PIRs}{\mbox{PIR}_s}
\newcommand{\tPI}{\widetilde{\mbox{PI}}}
\newcommand{\PI}{\mbox{PI}}
\newcommand{\PIs}{\mbox{PI}_s}
\newcommand{\PIz}{\mbox{MPI}}
\newcommand{\tPIz}{\widetilde{\mbox{MPI}}}
\newcommand{\PIzs}{\mbox{MPI}_s}
\newcommand{\V}{\mbox{VAR}}
\newcommand{\hV}{\widehat{\sigma}^2}
\newcommand{\tV}{\widetilde{\sigma}^2}

\newcommand{\tS}{\widetilde{\sigma}}
\newcommand{\hS}{\widehat{\sigma}}
\newcommand{\hmu}{\widehat{\mu}_Y}
\newcommand{\W}{\mbox{width}}
\newcommand{\bW}{\overline{\W}}
\newcommand{\hrho}{\widehat{\rho}}
\newcommand{\trho}{\widetilde{\rho}}
\newcommand{\hbeta}{\widehat{\beta}}

\begin{document}

\begin{center}

{\Large \bf On Correlation and Prediction Interval Reduction}

\bigskip

\textbf{Romain Piaget-Rossel \& Valentin Rousson}

\bigskip

\textit{Center for Primary Care and Public Health (Unisant\'e) \\ University of Lausanne, Switzerland}

\textit{Route de Berne 113, CH-1010 Lausanne, Switzerland}

\medskip

\texttt{romain.piaget-rossel@unisante.ch; valentin.rousson@unisante.ch}

\medskip

May 2024

\end{center}

\medskip

\noindent {\it Abstract}: Pearson's correlation coefficient is a popular statistical measure to summarize the strength of association between two continuous variables. It is usually interpreted via its square as percentage of variance of one variable predicted by the other in a linear regression model. It can be generalized for multiple regression via the coefficient of determination, which is not straightforward to interpret in terms of prediction accuracy. In this paper, we propose to assess the prediction accuracy of a linear model via the prediction interval reduction (PIR) by comparing the width of the prediction interval derived from this model with the width of the prediction interval obtained without this model. At the population level, PIR is one-to-one related to the correlation and the coefficient of determination. In particular, a correlation of 0.5 corresponds to a PIR of only 13\%. It is also the one's complement of the coefficient of alienation introduced at the beginning of last century. We argue that PIR is easily interpretable and useful to keep in mind how difficult it is to make accurate individual predictions, an important message in the era of precision medicine and artificial intelligence. Different estimates of PIR are compared in the context of a linear model and an extension of the PIR concept to non-linear models is outlined. 

\thispagestyle{empty}
\bigskip

\noindent {\it Key words}: coefficient of alienation; correlation; individual predictions; linear regression model; prediction accuracy; prediction interval.

\newpage

\section{Introduction}

Pearson's correlation coefficient (Galton, 1888; Pearson, 1895) is a famous statistical measure to summarize the strength of linear association between two (continuous) variables $X$ and $Y$, defined as the covariance between $X$ and $Y$ divided by the product of their standard deviations. Among other properties, it is invariant to affine transformations (i.e.\! it does not depend on the units of the variables) and it takes values between $-1$ and $+1$, the minimal and maximal values being attained in the case of a perfect linear relationship. While the sign of a correlation $\rho$ indicates the direction of the association, its amplitude is usually interpreted via its square $\rho^2$ as the percentage of the variance of one variable which can be predicted by the other using a least squares regression line. For example, a correlation of $\rho=\pm0.5$ implies that $\rho^2=25\%$ of the variance of one variable can be linearly predicted by the other, a correlation of $\rho=\pm1/\sqrt{2}=\pm0.71$ being necessary to reach $\rho^2=50\%$. This is what led Hull (1923) to write that ``the half of a perfect correlation is not 0.5 but 0.71''. 

The concept of $\rho^2$ can be generalized for multiple regression, where the goal is to predict an outcome $Y$ from $p$ predictors $X_1,\cdots,X_p$ via a multiple linear regression model:
\begin{equation}
Y=\beta_0+\beta_1X_1+\cdots+\beta_pX_p+\ve.
\label{mod}
\end{equation}
If the parameters $\beta_0,\beta_1,\cdots,\beta_p$ in (\ref{mod}) are estimated using least squares approach, the estimand $L=\beta_0+\beta_1X_1+\cdots+\beta_pX_p$ (or best linear predictor of $Y$) and the residual variable (or prediction error) $\ve=Y-L$ are uncorrelated, enabling a variance decomposition:
\begin{equation}
\V(Y)=\V(\beta_0+\beta_1X_1+\cdots+\beta_pX_p)+\V(\ve).
\end{equation}
Denoting the variances of the outcome and the residual variable respectively by $\sigma^2_Y$ and $\sigma^2_{\ve}$, a generalization of the square of Pearson correlation as percentage of variance of the outcome predicted by a linear model (\ref{mod}) can then be defined as: 
\begin{equation}
\rho^2=\frac{\V(\beta_0+\beta_1X_1+\cdots+\beta_pX_p)}{\V(Y)}
=\frac{\V(Y)-\V(\ve)}{\V(Y)}
=1-\frac{\sigma^2_{\ve}}{\sigma^2_Y}.
\label{rho2}
\end{equation}
This quantity $\rho^2$ is also called ``coefficient of determination'', which effectively coincides with the square of Pearson correlation when $p=1$.

The above interpretation of the square of a correlation is quite famous but not always appreciated by practitioners, as we argue below. In Section 2, we review some alternative attempts of interpretation of a correlation, in particular via the coefficient of alienation of Kelley (1919), or its one's complement. In Section 3, we propose a natural interpretation of the latter, which we call ``prediction interval reduction'' (PIR), and suggest its use to assess the prediction accuracy of a linear model. Different estimates of PIR are defined in Section 4, compared via simulations in Section 5, and illustrated with a real historical example in Section 6 . In Section 7, we sketch a possible generalization of the PIR concept to non-linear prediction models, such as those used in the modern algorithms of machine learning or artificial intelligence. Of note, our focus in this paper is purely prediction. We are not considering at all the delicate concepts of ``causality'' or of ``true model''. 

\section{Correlation and alienation}

The interpretation of the correlation has long been a matter of debate, one important issue being whether it should better be interpreted via its absolute or squared value (Tryon, 1929). Other transformations of the correlation have been proposed as summary measures of an association. If the correlation between an outcome and a predictor is given by $\rho$, Kelley (1919) noted that the correlation between the outcome and the induced residual variable, i.e.\! the quantity $\ve$ in the context of model (\ref{mod}), which he called ``the other factors'', is given by $\kappa=\sqrt{1-\rho^2}$. He called this quantity the ``coefficient of alienation'' and (considering $0<\rho<1$) noted that $\rho+\kappa>1$. For example, if the correlation between the outcome and the predictor reaches a value of $\rho=0.5$, the correlation between the outcome and ``the other factors'' will be as high as $\kappa=0.87$, underlying that ``it is eminently worthwhile to attempt to discover what they are'' (Kelley, 1919). 

By definition, $\rho^2+\kappa^2=1$, such that $\rho^2=\rho^2/(\rho^2+\kappa^2)$ can be interpreted as the percentage of the predicted variance, as recalled in Section 1. To avoid squaring while keeping a percentage, Nygaard (1926) suggested to measure prediction accuracy via $\rho/(\rho+\kappa)$, although his proposal proved to be ``extremely difficult to interpret'' (Tryon, 1919). On the other hand, Bingham (1931) suggested to use $1-\kappa$ in a context of a validity study, where the issue is to assess how close a test (e.g.\! to measure intelligence) is with respect to some ``criterion'' (or gold standard), the test being the predictor and the criterion the outcome. He called this  quantity ``coefficient of dependability'' and claimed that it ``gives an idea of the reliability of the unknown variable which is predicted from a given value of the known variable'', without being more explicit. In a textbook and considering a similar situation, Hull (1928, pp.\! 268--276) interpreted $1-\kappa$ as the ``percentage of forecasting efficiency'' and noted the ``remarkably small forecasting efficiencies corresponding to [correlation] values below 0.5''. Both authors provided tables enabling to move from $\rho$ to $1-\kappa$, similar to that reproduced below (see Table 1 in Section 3), which were however only rarely quoted.

Obviously, these coefficients $\kappa$ and $1-\kappa$ are related to the standard deviations of the outcome and of the residual variable in the context of model (\ref{mod}). For example, Nygaard (1926) noted that $\kappa$ is ``the ratio between the probable error when A is predicted from B and the probable error when A is merely selected as the average trait A'', while $1-\kappa$ is ``the percentage of reduction of the probable error when A is predicted from B from what it is when A is simply guessed to be the average trait A''. In other words, and whatever the number $p$ of predictors considered in (\ref{mod}), one has:
\begin{equation}
\kappa=\frac{\sigma_{\ve}}{\sigma_Y}
\end{equation}
or equivalently:
\begin{equation}
1-\kappa=1-\frac{ \sigma_{\ve}}{\sigma_Y}=\frac{\sigma_Y -\sigma_{\ve}}{\sigma_Y}. 
\end{equation}
Somewhat surprisingly, these considerations, which provide a useful interpretation of a correlation coefficient, have not been appreciated and exploited as much as they could/should in the literature. For example, Tryon (1919) noted that these concepts are ``difficult to grasp'', Nygaard (1926) was seeking for ``a method by which a much more direct and understandable interpretation may be made of the amount of relationship indicated by a coefficient of correlation'', while Brogden (1946) complained that both $\kappa$ and $1-\kappa$ ``are concerned with error of prediction and are not directly pertinent to the primary concern of prediction''. Such critics were unjustified and unfortunate, as argued in the next section. 

\begin{table}[h!]
\begin{center}
\begin{tabular}{|ccc|}
\hline
$\rho$ & $\rho^2$ & $\PIR$ \\
\hline
0 & 0 & 0\\
0.1 & 0.01 & 0.005  \\
0.2 & 0.04 & 0.02 \\
0.3 & 0.09 & 0.05  \\
0.4 & 0.16 & 0.08  \\
{\bf 0.5} & {\bf 0.25} & {\bf 0.13}  \\
0.6 & 0.36 & 0.2  \\
{\bf 0.707} & {\bf 0.5} & {\bf 0.29}  \\
 0.8 & 0.64 & 0.4\\
 {\bf 0.866} & {\bf 0.75} & {\bf 0.5} \\
 0.9 & 0.81 & 0.56\\
 0.95 & 0.90 & 0.69 \\
 0.99 & 0.98 & 0.86 \\
 0.995 & 0.99 & 0.9 \\
 0.999 & 0.998 & 0.955 \\
0.9999 & 0.9998 & 0.9859 \\
 1 & 1 & 1 \\
\hline
\end{tabular}
\caption{\textit{Relationship between a (positive) Pearson correlation $\rho$, the corresponding coefficient of determination $\rho^2$, and the prediction interval reduction $\PIR=1-\sqrt{1-\rho^2}$. In bold are the values where one of these three concepts reaches the value 0.5.}}
\end{center}
\end{table}

\section{Prediction interval reduction}

In the present paper, we follow Hull (1928) and Bingham (1931) and suggest to use the coefficient $1-\kappa$ (along with $\rho^2$) to routinely assess the prediction accuracy of a linear model, such as (\ref{mod}). As recalled above, $\rho^2$ can be interpreted as the percentage of the outcome's variance predicted by the linear model. Although it has nice mathematical properties, the variance is certainly not the most natural measure of variability in practice, as it expresses the variability in the squared units of the variable. This is why many practitioners will prefer the standard deviation over the variance to measure variability (e.g.\! Gaddis and Gaddis, 1990; Glass and Hopkins, 1996; McHugh, 2003).  For them, a percentage of predicted variance may not be a straightforward concept to assess prediction accuracy, e.g.\! when comparing several models in this regard. It is a bit as if someone wishing to calculate what the rent of an apartment represents in relation to a total budget did so in terms of squared dollars. Unfortunately, we cannot properly define something like ``a percentage of standard deviation of the outcome predicted by the linear model'' since the standard deviations of the best linear predictor and of the residual variable do not sum up to one. Nevertheless, we shall provide below the coefficient $1-\kappa$ with a useful interpretation, which will hopefully please the practitioners.

We have so far introduced model (\ref{mod}) without making any assumption, since writing $Y=L+(Y-L)$ is necessarily correct. Recall that $Y$ is the outcome while  the estimand $L=\beta_0+\beta_1X_1+\cdots+\beta_pX_p$ is the optimal linear combination of the $p$ predictors $X_1,\cdots,X_p$, where ``optimal'' is defined according to the least squares criterion. As recalled in Section 1, when least squares approach is used, it is a mathematical fact that $L$ and the residual variable $\ve=Y-L$ are uncorrelated (and it is also a mathematical fact that $\ve$ has mean zero). We shall now assume that $L$ and $\ve$ are not only uncorrelated but independent, i.e.\! that the conditional distribution of the residual variable given the values of the predictors is the same whatever these values. As done in a classical linear regression model, we shall assume that this unique conditional distribution is normal. Equivalently, we shall assume that the conditional distribution of the outcome $Y$ given the values of the predictors $X_1=x_1,\cdots,X_p=x_p$ is (1) normal (normality assumption), (2) with mean $\beta_0+\beta_1x_1+\cdots+\beta_px_p=\mathbf{x}\bm{\beta}$, where $\mathbf{x}=(1,x_1,\cdots,x_p)$ and $\bm{\beta}=(\beta_0,\beta_1,\cdots,\beta_p)'$ (linearity assumption), and (3) with standard deviation $\sigma_{\ve}$, the variability being the same whatever the values of the predictors (homoscedasticity assumption). In what follows, we refer to this set of three assumptions as the ``linear model assumptions". 

Importantly, if the goal is to predict the outcome $Y$ for an individual with $X_1=x_1,\cdots,X_p=x_p$, it is not sufficient to just provide a point prediction; 
one should also provide a prediction interval constructed around that point prediction. A prediction interval is said to be valid if it contains the true outcome value with a high given probability (level), typically 95\%. Thus, the width of a prediction interval informs us about the accuracy of the prediction. If the prediction is done via a linear model (\ref{mod}) with $p$ predictors, and if we make the linear model assumptions, a valid prediction interval at the level $\gamma$ is obtained by: 
\begin{equation}
\PI(\mathbf{x})=
\mathbf{x}\bm{\beta}\pm z_{(1+\gamma)/2} \cdot \sigma_{\ve}.
\label{pi1}
\end{equation}    
In this formula, $z_{(1+\gamma)/2}$ denotes the $(1+\gamma)/2$ quantile of the standard normal distribution (e.g.\! $z_{(1+\gamma)/2}=1.96$ if $\gamma=95\%$). The width of $\PI(\mathbf{x})$ is thus independent of $\mathbf{x}$ and given by $\W(\PI(\mathbf{x}))=2z_{(1+\gamma)/2} \cdot \sigma_{\ve}$.

To measure the prediction accuracy achieved with model (\ref{mod}), we propose to compare the width of $\PI(\mathbf{x})$ defined in (\ref{pi1}) with the width of another prediction interval calculated without model (\ref{mod}), in a situation where the prediction of the outcome for an individual is made without knowing any characteristics of that individual, and where one would just assume a marginal normal distribution for the outcome. Noting by $\mu_Y$ and $\sigma_Y$ the mean and standard deviation of the outcome $Y$, a valid marginal prediction interval at the level  $\gamma$ is then obtained by: 
\begin{equation}
\PIz=\mu_Y \pm z_{(1+\gamma)/2} \cdot  \sigma_Y.
\label{pi0}
\end{equation}    
The width of $\PIz$ is given by $\W(\PIz)=2z_{(1+\gamma)/2} \cdot \sigma_Y$. Of course, $\PI(\mathbf{x})$ should be narrower than $\PIz$, the question being by how much. As done in Rousson's French statistical textbook (2013, Chapter 13), also inspired by a manuscript by Gasser and Seifert (2002, p.\! 93), we propose to measure the prediction accuracy of model (\ref{mod}) as the percentage width reduction of $\PI(\mathbf{x})$ as compared to $\PIz$. Such a ``prediction interval reduction'' (PIR) is defined by:
\begin{equation}
\PIR=\frac{\W(\PIz)-\W(\PI(\mathbf{x}))}{\W(\PIz)}.
\end{equation}
Noticeably, one also has:
\begin{equation}
\PIR
=\frac{\sigma_Y-\sigma_{\ve}}{\sigma_Y}
=1-\kappa=1-\sqrt{1-\rho^2}.
\end{equation}
The $\PIR$ coefficient is thus equivalent to the coefficient $1-\kappa$ above, at least asymptotically, i.e.\! at the population level or in large samples. In small samples, these two concepts will differ a bit, as discussed in the next two sections. Table 1 provides examples on how to move from $\rho$ (or $\rho^2$) to $\PIR$. For example, a correlation of $\rho=\pm0.5$ between an outcome and a predictor implies that the width of the prediction interval for the outcome of an individual with a given value for that predictor would be reduced (only) by a factor $\PIR=13\%$ compared to a prediction interval calculated without any predictor, a correlation of $\rho=\pm\sqrt{0.75}=\pm0.87$ being necessary to reach $\PIR=50\%$. We shall thus amend the aforementioned Hull's statement and write instead that ``the half of a perfect correlation is not 0.5 but 0.87'', which has the merit to emphasize the difficulty of making individual predictions. 

\section{Estimation}

The concept of prediction interval reduction presented in the last section was introduced at the population level. We discuss here how this can be calculated from a sample of data. We consider a sample of $n$ multivariate and independent observations of $(X_1,\cdots,X_p,Y)$ assuming the linear model assumptions in (\ref{mod}). Let $\mathbf{y}$ be the $n$-dimensional vector, such that $y_i$ is the measurement of variable $Y$ for observation $i$, and let $\mathbf{X}$ be the $n\times(p+1)$ ``design matrix'' with elements $x_{ij}$, such that $x_{i1}=1$ and $x_{ij}$ is the measurement of variable $X_{j-1}$ for observation $i$. Let $\bm{\hbeta}=(\mathbf{X}'\mathbf{X})^{-1}X'\mathbf{y}=(\hbeta_0,\hbeta_1,\cdots,\hbeta_p)'$ be the least-squares estimate of the vector of parameters $\bm{\beta}$. Let $\bm{\heps}=\mathbf{y}-\mathbf{X}\bm{\hbeta}$ be the vector of  sample residuals. Let $\hV_{\ve}=\bm{\heps}'\bm{\heps}/n$ and $\tV_{\ve}=\bm{\heps}'\bm{\heps}/(n-p-1)$ be respectively a biased and an unbiased estimate of the residual variance $\sigma^2_{\ve}$, and let $\hV_Y=(\mathbf{y}-\hmu \mathbf{1})'(\mathbf{y}-\hmu \mathbf{1})/n$ and $\tV_Y=(\mathbf{y}-\hmu \mathbf{1})'(\mathbf{y}-\hmu \mathbf{1})/(n-1)$ be respectively a biased and an unbiased estimate of the outcome variance $\sigma^2_Y$, where $\hmu$ denotes the sample mean of the $y_i$ and $\mathbf{1}$ is an $n$-dimensional vector of ones. 

Prediction interval (\ref{pi1}) at the level $\gamma$ for the outcome $Y$ given the values of the predictors $X_1=x_1,\cdots,X_p=x_p$ can be estimated by:
\begin{equation}
\tPI(\mathbf{x})=\mathbf{x}\bm{\hbeta}  \pm z_{(1+\gamma)/2} \cdot \tS_{\ve}.
\label{hpi1a}
\end{equation}
Under the linear model assumptions, however, an exact prediction interval taking into account  the uncertainty of the estimates is also available as (e.g.\! Montgomery et al., 2012):
\begin{equation}
\PIs(\mathbf{x})=\mathbf{x}\bm{\hbeta}  \pm t_{(1+\gamma)/2,n-p-1} \cdot \tS_{\ve} \cdot \sqrt{1+\mathbf{x}(\mathbf{X}'\mathbf{X})^{-1}\mathbf{x}'}.
\label{hpi1}
\end{equation}
In this formula, $t_{(1+\gamma)/2,n-p-1}$ denotes the $(1+\gamma)/2$ quantile of a Student distribution with $n-p-1$ degrees of freedom. Contrary to (\ref{hpi1a}), prediction interval (\ref{hpi1}) will become (slightly) wider as $\mathbf{x}$ gets away from the center of gravity of the data in the space of the predictors, although (\ref{hpi1a}) and (\ref{hpi1}) will almost coincide for a large $n$ (compared to $p$). 

In analogy with (\ref{hpi1a}), a marginal prediction interval (\ref{pi0}) at the level $\gamma$ for the outcome $Y$ can be estimated by:
\begin{equation}
\tPIz=\hmu   \pm z_{(1+\gamma)/2} \cdot \tS_Y.
\label{hpi0a}
\end{equation}
Here also, under the marginal normal assumption of the outcome, an exact formula is available as:
\begin{equation}
\PIzs=\hmu   \pm t_{(1+\gamma)/2,n-1} \cdot \tS_Y\cdot \sqrt{1+1/n}.
\label{hpi0}
\end{equation}
Thus, a sample version of PIR is given by:
\begin{equation}
\PIRs=\frac{\W(\PIzs)-\bW(\PIs(\mathbf{x}))}{\W(\PIzs)}.
\end{equation}
In this formula, $\bW(\PIs(\mathbf{x}))$ denotes the average width of the $\PIs(\mathbf{x})$ calculated over the vectors of predictors $\mathbf{x}=\mathbf{x}_i$ corresponding to the $n$ observations in the sample, where $\mathbf{x}_i=(1,x_{i1},\cdots,x_{ip})$. This quantity $\PIRs$ can be considered as an estimate of PIR defined in the population. On the other hand, it can also be considered as the quantity of interest since it informs us about the prediction interval reduction obtained in practice, taking into account that we do not know the true parameters in (\ref{mod}), while PIR is the prediction interval reduction that would be obtained theoretically if the true parameters were known.

To estimate PIR in a simpler way, one could alternatively use the following two quantities: 
\begin{equation}
\trho^2=1-\frac{\tV_{\ve}}{\tV_Y}
\quad \mbox{and} \quad 
\hrho^2=1-\frac{\hV_{\ve}}{\hV_Y}.
\label{R2}
\end{equation}
These two quantities, often denoted respectively as $R^2_{adj}$ and $R^2$ in the statistical literature, are two consistent estimates of the coefficient of determination $\rho^2$ when $n$ is large (compared to $p$). While the former can also be used when $n$ is small (compared to $p$), the latter is the square of the sample Pearson correlation between the predictor and the outcome when $p=1$. Two natural estimates of PIR are then given by: 
\begin{equation}
\tPIR=1-\sqrt{1-\trho^2}=\frac{\tS_Y-\tS_{\ve}}{\tS_Y}=\frac{\W(\tPIz)-\W(\tPI(\mathbf{x}))}{\W(\tPIz)}
\end{equation}
and by:
\begin{equation}
\hPIR=1-\sqrt{1-\hrho^2}=\frac{\hS_Y-\hS_{\ve}}{\hS_Y}.
\label{PIRh}
\end{equation}
Estimates $\tPIR$ and $\hPIR$ will be compared with $\PIRs$ via simulations in the next section.
 
\section{Simulations}

In this section, we present the results of simulations to compare the different estimates of PIR presented in the previous section. Multivariate samples with $n=20,100$ or 500 observations of $(X_1,\cdots,X_p,Y)$  were simulated according to model (\ref{mod}) under the linear model assumptions, with $\beta_0=\beta_1=\cdots=\beta_p=1$. We considered $p=1, 5$ or $10$ predictor(s) following a univariate (in case $p=1$) or a multivariate (in case $p>1$) normal distribution where each component had mean 0 and variance 1, and where correlations between components were equal to $\rho_x=0$ or 0.5. The residual variable $\ve$ in (\ref{mod}) was taken to be normally distributed with mean 0 and with variance $\sigma^2_{\ve}$ chosen such that the coefficient of determination was equal to $\rho^2=0.25, 0.5$ or 0.9, that is:
\begin{equation}
\sigma^2_{\ve}=\frac{1-\rho^2}{\rho^2}\cdotp(1+(p-1)\rho_x).
\end{equation}
In each sample hence generated, estimates $\PIRs$, $\hPIR$ and $\tPIR$ have been calculated. Recall that at the population level, we have $\PIR=1-\sqrt{1-\rho^2}$. 

Figure 1 shows boxplots summarizing the results obtained from 2000 simulations in all settings considered with $\rho_x=0$. Results were similar with $\rho_x=0.5$ illustrating the fact that our results did not depend on the distribution of the predictors. Clearly, the three estimates converged towards PIR as $n$ increased, being almost identical to each other when $n=500$ whatever the value of $p$ or $\rho^2$. For smaller values of $n$, $\tPIR$ was still an approximately unbiased estimate of PIR, whereas $\hPIR$ had an upward bias and $\PIRs$ had a downward bias with respect to PIR, especially for a large $p$. This is because $\PIRs$ is based on exact prediction intervals, which takes into account the uncertainty of the estimates and thus becomes wider as the number of parameters to estimate increases. For $p=1$, however, the three estimates did not differ much from each other, even with $n=20$. 

To sum up, estimate $\tPIR$, which is directly obtainable from the adjusted coefficient of determination $\trho^2=R^2_{adj}$, can be regarded as an approximately unbiased estimate of PIR, the prediction interval reduction that would be achievable at the population level. In this regard, it is a better estimate than $\hPIR$, which is directly obtainable from the (unadjusted) coefficient of determination $\hrho^2=R^2$, although negligibly so for $p=1$, where $\hrho^2$ is simply the square of the sample Pearson correlation. On the other hand, $\PIRs$ informs us about the prediction interval reduction achievable using our sample estimates, which tends to be smaller than PIR, although negligibly so for a large sample size. 

\section{Example}

We illustrate the PIR concept using real historical data collected by Karl Pearson himself, the {\tt father.son} data set, which can be found in the {\tt UsingR} library from the statistical package R (Verzani, 2022). These data consist of the height of the father and the height of one of his fully grown son for $n=1078$ pairs of father and son collected in England at the end of the 19th century. These data (here expressed in cm) are plotted on the left panel of Figure 2. We consider that the goal is to predict the height of the son ($Y$) knowing the height of his father ($X$). Remarkably, Pearson's correlation coefficient between $X$ and $Y$ estimated from these data is precisely $\hrho=0.50$ (when rounded to two decimals), the percentage of the variance of  son's height which is linearly predicted by father's height being estimated to $\hrho^2=R^2=25\%$. Note that due to the large sample size, we also have here (when rounded to two decimals) $\trho^2=R^2_{adj}=25\%$.  

As can be seen from the left panel of Figure 2, and as it is often the case with height measurements (e.g.\! A'Hearn et al., 2009), the data look pretty close to a binormal distribution. As a consequence, the ``linear model assumptions'' are here reasonable and valid prediction intervals can be calculated. The intercept and slope of a linear model (with a unique predictor) estimated from these data are respectively $\hbeta_0=86.07$ and $\hbeta_1=0.51$, while the estimated residual standard deviation is $\tS_{\ve}=6.19$. Thus, a prediction interval at the 95\% level for a son's height knowing his father's height $x$ can be estimated via (\ref{hpi1a}) by $\tPI(x)= 86.07+0.51x\pm1.96\cdot 6.19$. For instance, one gets\! [156.2;180.5] with $x=160$, and one gets [166.5;190.7] with $x=180$, the latter containing logically higher values than the former, although the two prediction intervals largely overlap, as illustrated in the left panel of Figure 2, where other examples are also shown. Note that the prediction intervals shown in Figure 2 were calculated using the exact formula (\ref{hpi1}), which provides almost the same results as those obtained with the approximate formula (\ref{hpi1a}). To check their validity, we note that 1028 out of the 1078 son's height (95.4\%) in this data set truly lie within the prediction intervals calculated from their respective father's height, which is remarkably close to the theoretical 95\%. As consequences of the normality and homoscedasticity assumptions, all those prediction intervals (calculated with different values of $x$) share (approximately) the same width, given by $2\cdot1.96\cdot6.19=24.3$. In other words, the uncertainty of a prediction obtained here is $\pm12.15$ cm. In what follows, we shall compare this uncertainty with the uncertainty obtained when making a prediction for a son's height without knowing his father's height.

At first sight, it might seem impossible to predict the height of someone without knowing anything about him. Yet, since we know a bit the concept of height, we are able to exclude some impossible (or very unlikely) values, such as 20 or 300 cm. Assuming marginal normality of the distribution of height, it is even possible to calculate a valid prediction interval by just estimating its mean and standard deviation. Using our data, where we implicitly consider the son's height distribution in England at the end of the 19th century, we obtain $\hmu=174.5$ and $\tS_Y=7.15$. Thus, a marginal prediction interval at the 95\% level for a son's height without knowing anything about him can be estimated via (\ref{hpi0a}) by $\tPIz=174.5\pm1.96\cdot7.15=[160.4;188.5]$, as illustrated on the right panel of Figure 2. Here also, one can check its validity by noting that 1023 out of the 1078 son's height (94.9\%) in this data set truly lie within that prediction interval (again extremely close to the theoretical 95\%). The width of such a prediction interval is  $2\cdot1.96\cdot7.15=28.0$. The uncertainty of a prediction for a son's height made without any information about him (except that he belongs to the male population in England at the end of the 19th century) is therefore $\pm14.0$ cm, representing the ``basic uncertainty''. To sum up, when knowing and exploiting father's height to predict a son's height, we are able to reduce the uncertainty of the prediction from a basic uncertainty of $\pm14.0$ cm to $\pm 12.15$ cm. Expressed as percentage of the basic uncertainty, this reduction amounts to $(14.0-12.15)/14.0=13\%$, which is precisely an estimation of PIR. It could equivalently be estimated by comparing $\tS_Y=7.15$ and $\tS_{\ve}=6.19$, or by transforming the sample Pearson correlation $\hrho=0.50$ as follows:
\begin{equation}
\tPIR=\frac{7.15-6.19}{7.15}=13\% 
\quad\mbox{or}\quad 
\hPIR=1-\sqrt{1-0.50^2}=13\%.
\end{equation}
Note that due to the large sample size, the three estimates of PIR considered above are here almost identical since we have $\PIRs=0.1339$, $\tPIR=0.1343$ and $\hPIR=0.1347$.
This example is a nice illustration that a correlation of 0.5, and thus a percentage of predicted variance of  25\%, corresponds to a prediction interval reduction of (only) 13\%.

\section{Extensions and conclusion}

In this paper, we have provided a useful interpretation of (the one's complement of) the coefficient of alienation as the prediction interval reduction (PIR), and we propose its use (along with the coefficient of determination) to assess the prediction accuracy of a linear model. The PIR coefficient informs us about the reduction of uncertainty (defined as the width of a prediction interval) achieved using a linear model, compared to the ``basic uncertainty'' which would be achieved without any model, expressed as the percentage of the basic uncertainty. It also shares with the coefficient of determination the property of being a percentage, so one ought not to be an expert in the field to get an idea about the prediction accuracy of a model. Finally, since a PIR coefficient is appreciably lower than a correlation or a coefficient of determination, it has the merit to remind us that making individual predictions is an extremely difficult task, a reality which is sometimes forgotten in the era of precision medicine and artificial intelligence. 

The above interpretation of the PIR coefficient is based on its definition: 
\begin{equation}
\PIR=\frac{\W(\PIz)-\W(\PI(\mathbf{x}))}{\W(\PIz)}.
\label{gen}
\end{equation}
It is related to Pearson's correlation coefficient and the coefficient of determination under (I) the classical assumptions of a linear model (normality, linearity, homoscedasticity), which are necessary to calculate $\PI(\mathbf{x})$, and (II) the marginal normality assumption of the outcome which is necessary to calculate $\PIz$. Note that both (I) and (II) are  satisfied in the case of a joint multinormal distribution of the outcome and the predictors, whereas under (I) with non-normal predictors, (II) will still approximately hold in the case of a weak association between the outcome and the predictors. 
The PIR concept could also be extended to cases where (I) and (II) are not satisfied (although it would no longer be related to Pearson correlation and the coefficient of determination). In particular, it is not difficult to calculate a valid marginal prediction interval $\PIz$ when (II) is not satisfied. One could just use appropriate quantiles of the outcome distribution to define the lower and the upper bounds of $\PIz$ (e.g.\! the 2.5\% and 97.5\% quantiles to get a prediction interval at the level 95\%), which can be empirically (non-parametrically) estimated from the data. It is somewhat more difficult to calculate a valid $\PI(\mathbf{x})$ when (I) is not satisfied, e.g.\! in the context of non-linear models, such as those used in machine learning. If the normality and homoscedasticity assumptions are violated, an additional difficulty is that the width of $\PI(\mathbf{x})$ may strongly depend on the individual for whom (i.e.\! the value of $\mathbf{x}$ for which) it is calculated. One solution would be to replace the unique width of $\PI(\mathbf{x})$ in the definition of $\PIR$ with an average width of the $\PI(\mathbf{x})$ calculated over all (or some) individuals in the sample, noted $\bW(\PI(\mathbf{x}))$, yielding the following generalization of the PIR concept: 
\begin{equation}
\PIR=\frac{\W(\PIz)-\bW(\PI(\mathbf{x}))}{\W(\PIz)}.
\end{equation}
The calculation of prediction intervals might be non-trivial in some non-parametric or non-explicit models, although possible via simulations (e.g.\! Khosravi et al., 2011). As recalled in Section 3, having just a point prediction without any idea about its accuracy is almost useless, such that the calculation of valid prediction intervals should actually be a prerequisite for any method or model used for making individual predictions. A sensible generalization of PIR to binary or categorical outcomes would also be welcome, although the concept of prediction interval might be less natural for discrete outcomes.

To conclude, it is beneficial to all of us (scientists, politicians, journalists, citizens) to be able to properly assess and appreciate scientific progress, as in medicine or the social sciences, in domains where one would like to rely on the quality of individual predictions. How good are the current predictions? Are we above or below PIR values of 50\% in a given field? What do we want to attain in the future? Are we aiming for 90\%? Or even more? Debating on what is realistic to achieve given the current state of science is a challenging and useful question, and regular use of the PIR concept could help in this regard. It will also be helpful for students and researchers to realize that correlations as high as 0.87 represent an uncertainty reduction of only 50\%, reminding us of about the existence of the many unknown factors and the underlying difficulties of the scientific task.

\medskip
\medskip

\noindent {\bf Disclosure statement}: The authors report no conflict of interest.

\begin{figure}[h!] 
\begin{center}
\hspace*{-1cm}
\includegraphics[scale=0.8]{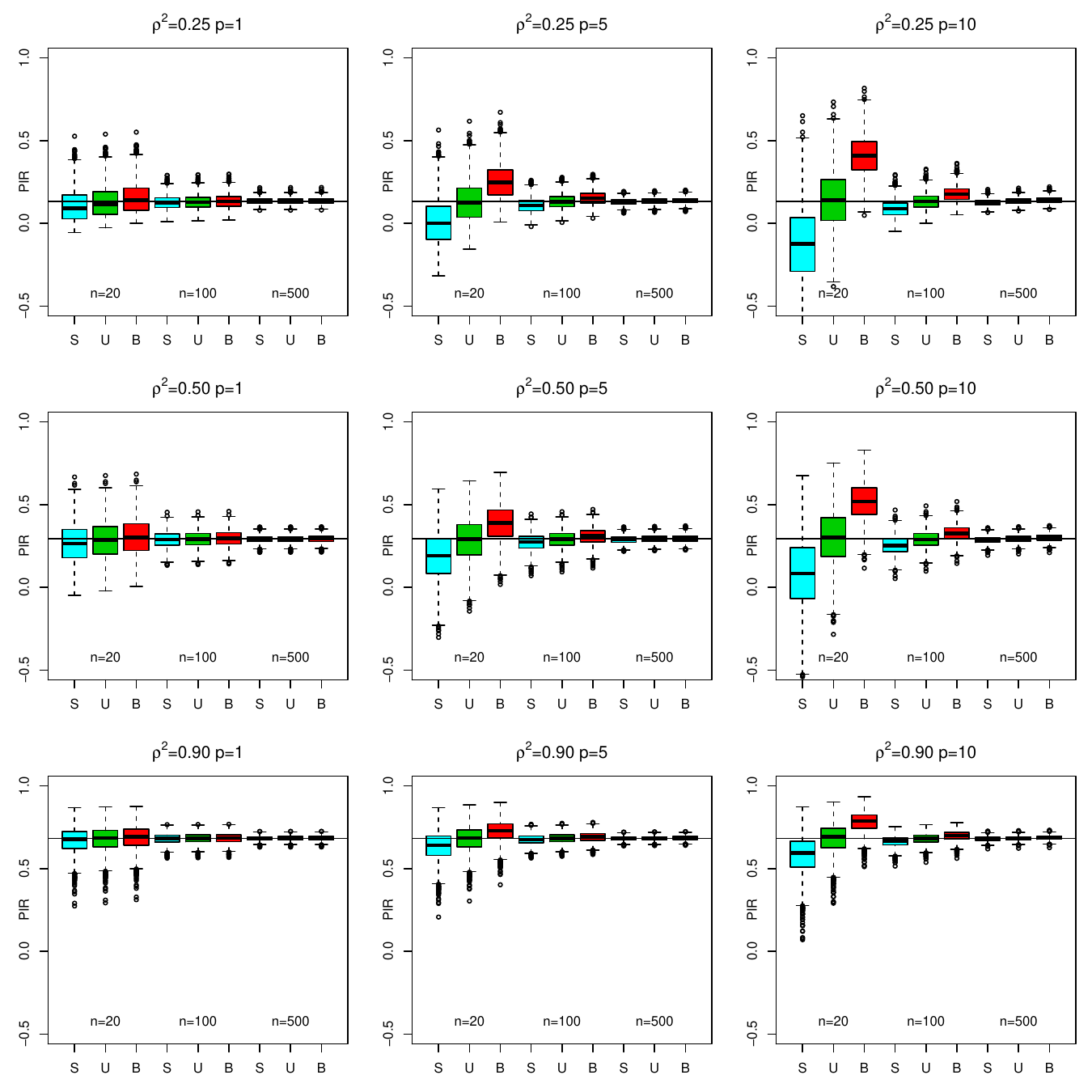}
\end{center}
\caption{\textit{Boxplots of 2000 estimates $\PIRs$ (labelled with an `S' for sample), $\tPIR$ (labelled with an `U' for unbiased) and $\hPIR$ (labelled with a `B' for biased) calculated from a sample of $n=20, 100$ or 500 observations simulated from regression model (\ref{mod}) with $p=1, 5$ or 10 predictors and a coefficient of determination of $\rho^2=0.25, 0.50$ or 0.90. In each panel, the horizontal line indicates the value of $\PIR=1-\sqrt{1-\rho^2}$ in the population.}}
\end{figure}

\begin{figure}[h!]
\begin{center}
\hspace*{-1cm}
\includegraphics[scale=1]{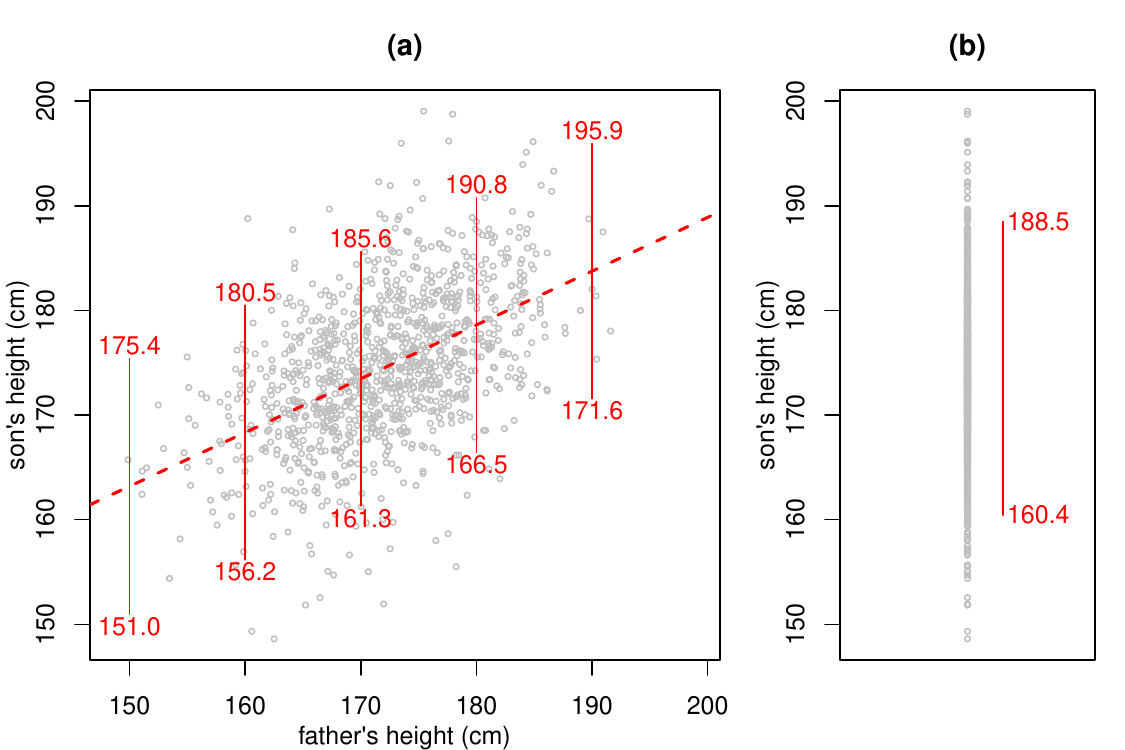}
\end{center}
\caption{\textit{Vertical lines represent prediction intervals at the level 95\% for a son's height, calculated (a) knowing or (b) without knowing his father's height, from height data of $n=1078$ couples of father and son (grey dots). The dotted line on panel (a) represents the least squares regression line, defining the centers of the prediction intervals.}}
\end{figure}

\end{document}